\documentstyle[12pt]{article}
\input psfig
\long\def\comment#1{}
\begin{document}
\title{Quantum Information and Entropy}
\author{Subhash Kak}
\date{}
\maketitle

\begin{abstract}

Thermodynamic 
entropy is not an
entirely satisfactory  measure of information of a quantum
state. This entropy for an
unknown pure state is zero, although
repeated measurements on copies of 
such a pure state do communicate information.
In view of this, we propose a new measure for the
informational entropy of a quantum state that includes
information in the pure states and 
the thermodynamic entropy.
The origin of information is explained in terms of an
interplay between unitary and non-unitary evolution.
Such complementarity is also at the basis of the
so-called interaction-free measurement.

\end{abstract}

\section{Introduction}

Thermodynamic entropy measures the disorder of a system, and although
we will show that 
 it is not
identical to informational entropy, it is used 
freely in physics
and employed in settings
where not only order but also what is intuitively ``information'' are involved.
In the popular
view that information
is the foundational stuff of reality, what is
meant is informational
entropy, but what is used is thermodynamic entropy.

Thermodynamic entropy considers 
the number of structural arrangements associated
with the system, whereas informational entropy
is about choices made in a communications context. The argument might be
made that information is ultimately physical and, therefore, there should
be a thermodynamic basis to informational entropy. But this is true only as 
long as it is possible to characterize the information process in 
terms of statistical ensembles, which may not be the case in situations 
relating to communicating agents or in 
quantum cosmology.

In a classical system,
informational entropy
may be best viewed in the context of a game between
the source, {\it X}, and the receiver, {\it Y}, in which, upon
receipt of signal, the receiver discovers which
signal was actually sent 
(Here we don't concern ourselves with complications arising
out of noisy communication).
The source chooses a signal out of an ensemble, and the choices
are repeated in accord with the language (patterns of
signals) that connects
it with the receiver.
For physical systems, the game may be perceived as being played
between Nature and the physicist.

The same idea of the game also underlies quantum information [1]. But
here the situation is more complex, because the quantum state could
be pure or mixed, and these two cases are very different from the
point of view of measurement.
A mixed state is a statistical mixture of component pure states,
and its entropy is computed by the von Neumann measure in a manner
that is similar to the entropy for classical states.
A pure state is completely described by its state function and its
von Neumann entropy is zero.

It is important to note the asymmetry between the quantum system
and the physicist.
From the point of view of the preparer of the states, the pure state
carries information that is limited by the ``relationship''
between the source and the receiver, and by the precision of the
receiver's measurement
apparatus. The source may choose out of an infinity of possibilities,
and the dependence on the ``relationship'' implies that the
pure state's information will vary from one receiver to another.

For the source, the information generated by him equals the probability
of choosing the specific state out of the possibilities available to
him (this is the states {\it a priori} probability).
If the set of choices is infinite, then the ``information'' generated by
the source is unbounded.
On the other hand, due to the probabilistic nature of the reception
process, not 
all the information at the source is obtained
at the receiver by his measurement.

In recent years several theories have been advanced that assign
finite entropy to matter and space [2].
The finite value of entropy for a given volume has been taken to mean that
matter cannot be subdivided infinitely, and that the 
fundamental entity relating to matter
is a bit (1 or 0) of information.

However, this approach of discretization
hasn't been very successful. Part of the fault may lie in 
the limitations of the current concept of quantum entropy.
In particular, von Neumann entropy is not the right measure in
the asymmetric situation where the choice of the state itself carries
information.

In this paper, we argue that an unknown quantum pure state, when viewed
in the context of the game between the source and the receiver, communicates
information just as a mixed
state. 
We propose a measure for informational entropy and show
that it may be seen as the sum of information
in the pure states and the thermodynamic entropy. 
The origin of information is seen as a consequence
of the interplay between unitary and non-unitary evolution,
which makes it possible to transform one type of information
into another.
The significance of this complementarity is considered for
the case of ``interaction-free'' measurements.
This complementarity indicates that a fundamental
duality is essential for information, which means that complete unification 
will not be possible.

\section{Classical and von Neumann measures of information}

Let the source be associated
with a random variable, X, that takes values from a discrete set
$x_1 , x_2 , ...,  x_n$ with probabilities $p(x_1 ), p(x_2 ),..., p(x_n )$.
The information associated with the receipt of signal $x_i $ is
$-\log_2 p(x_i )$. The average information, or Shannon entropy, of the
source is:

\begin{equation}
 H(X) = - \sum_i p(x_i ) \log p (x_i ).
\end{equation}

The maximum value of entropy, obtained for the case when
all signals are equally likely, is $\log n$.
When the variable $X$ is continuous
with the probability density $f_X (x)$,
its entropy $H(X)$
is given by the expression:

\begin{equation}
H (X) = h(X) - \lim_{\Delta x \rightarrow 0} \log \Delta x,
\end{equation}

\noindent
where $h(X)$ is the Boltzmann or differential entropy:

\begin{equation}
h(X) = \int_{- \infty}^{\infty} f_X (x) \log \left[\frac{1}{f_X (x)}\right] dx,
\end{equation}

\noindent
and $\Delta x$ is the precision associated with the 
measurement of the variable.
The value of $H(X)$  
depends on the details of the experimental
arrangement.
Its maximum value, when the precision is absolute, is infinite.

If it is taken that the measurement has its own uncertainty, then
the value of entropy is finite that is given by the well-known 
information capacity theorem.

The measure of entropy (1), when generalized for a quantum system 
characterized by the density operator $\rho$,
is the von Neumann
entropy:

\begin{equation}
 S_n (\rho) = - tr(\rho \log \rho),
\end{equation}

\noindent
This may be equivalently written as:
\begin{equation}
 S_n (\rho) = - \sum_x \lambda_x \log \lambda_x, 
\end{equation}

\noindent
where  $\lambda_x$ are the eigenvalues of the density matrix $\rho $ associated
with the system.

The von Neumann entropy may be viewed as the average information the
experimenter obtains in the repeated observations of the very many
copies of an identically prepared mixed state. 
The entropy $S(\rho)$ for the mixed state

\begin{equation}
\rho  =   \left[ \begin{array}{cc}
                                  p  & 0 \\
                                  0 & 1-p \\
                               \end{array} \right]
\end{equation}

\noindent
is equal to 
\begin{equation}
-p \log p - (1-p) \log (1-p) .
\end{equation}

The von Neumann entropy of a pure state is zero, indicating that once it
has been identified then there is no further information to
be obtained from its copies, which is not the case with a 
mixed state.

\section{Entropy of the universe}

When applied to matter, some general arguments related
to degrees of freedom are
invoked to  estimate that
the entropy of a physical system is equal to

\begin{equation}
S_n \leq \frac{A}{4} 
\end{equation}

\noindent
where A is the area in Planck units equal to $\hbar G/c^3$.
This is the Bekenstein bound [2],
given originally in the form
$S \leq 2 \pi E L$ , where L is the linear size of the region, and E is the 
energy.
Gerard 't Hooft later generalized it
[3] to the form involving A/4 nats (1 bit
equals $\ln 2$ nats), and as
the holographic principle [4] it is supposed 
to apply to all matter.
The total quantity of bits in this approach is a measure of the 
degrees of freedom associated with the system. 
An informational approach based on fundamental limitation to precision 
of the measurement also indicates finite entropy. But if such a 
limitation is not justified, then information associated with space and
matter should be infinite.

A physical system is
described in terms of its state at some specific time, and the dynamical laws
governing its evolution. 
The idea of entropy tells us which configurations are more
likely than others. 
For dynamical laws, one expects dimensionless parameters in a
theory to be of order unity, reflecting the interaction 
between comparable processes.
But the gravitational, weak, and strong forces have 
characteristic dimensions
that are of very different orders of magnitude.
Furthermore, the actual range spanned by parameters related to gravitation,
electro-weak and strong forces, and the Hubble scale characteristic 
of cosmology is
immense, indicating that we may not be looking at the question the 
correct way.
 
The distribution of matter on very large scales has been found to be 
approximately
homogeneous and isotropic. The current data is interpreted to mean that distant
galaxies are expanding away from each other in accordance with Hubble's law. 
 
By extrapolation into the past, the universe is taken to have 
originated about 14
billion years ago in a superdense state. 
If it was a state in thermal equilibrium, then this would 
mean a violation of the
second law of thermodynamics, since the initial state should be in
an entropy minimum.

In the current synthesis, ``ordinary matter,''
consisting of particles
described by the Standard Model of particle physics, accounts for only about 
4\% of the
total energy of the universe. 
It is believed that another 23\% comes from particles yet
to be discovered, or ``dark matter,'' and a further 73\% 
is ``dark energy,''
generated by an unknown force.
 
The matter in the universe appears to be smoothly distributed, and the 
deviations
from smoothness are taken as a consequence of initial conditions [5]. 
The entropy of
matter and radiation in the observable universe is approximately 
$10^{88}$, where it is assumed that the background
radiation entropy for each baryon is $10^8$. 
(The entropy here is in ``natural units,'' in which the Boltzmann's
constant is taken to be unity.)
Initially this was mostly in the form of radiation, but now
it is assumed to be mainly concentrated in the entropy of the black holes at the
centers of galaxies.
 
With probably more than ten billion galaxies with million-solar-mass black holes at
their centers, the current entropy in black holes is of the order of $10^{100}$. If all the
matter in the observable universe were to be combined into one giant black hole, the
entropy would be significantly larger, $10^{120}$. 
 
The estimated entropy of the universe is rather small, given 
the size of the universe. Although it is 
believed to be
increasing due to the second law of thermodynamics, but it is lagging
its potential maximum. 
Others have argued that the initial entropy must have been still lower, with
estimates from $10^{10}$ to $10^{20}$.

The von Neumann measure leads to the puzzle of how information
arose in the universe. If we were to assume that the total universe
quantum state in the beginning was pure, then the information associated
with the universe as a whole was zero.
On the other hand, if it is assumed that the
deviations from perfect isotropy represented the
initial entropy, then the amount of
this entropy was rather small.
If the components now are entangled states, their ancestor states
at the beginning should also have been entangled.

A related puzzle is the emergence of non-unitary evolution in
the universe.
There can be
no information in a universe completely governed by unitary 
evolution. The resolution to 
this puzzle is to assume that the  physical universe comes with evolution
that has unitary as well as non-unitary components.
This duality is what makes information possible in the
universe.
It follows that one cannot assume a single mechanism 
behind the two evolutions.

The question of  how information is increasing is a central one
in physics.
Our proposed measure provides a resolution by showing how pure states 
carry entropy.

\section{Informational entropy, $S_i (\rho)$, of a quantum system}

Unlike a classical state that is completely known when it is measured,
the process of measurement of
a quantum state merely determines its projection along chosen
basis vectors, and this projection is probabilistic.

Once a classical variable has been measured (examples being location
or mass), it is correct to assume that further measurements will
not provide any new information.
In the case of location variable, we know that the object will
continue at its position owing to the fact that the object
can, in principle, be isolated from the environment.
Likewise, the mass values, in further measurements, will be
identical to the first measurement.

Let the setting for the game related to quantum information be
one where the source is producing identical copies
of an elementary quantum state. In contrast to the classical case,
there are two different situations that one must consider.
In general, one doesn't know whether the state is pure or mixed.
The game for the receiver is to determine this state as closely
as possible, after examining as many copies of the state as is required.
The entropy then is the average information communicated about the unknown 
state at any point in the measurement process.

It is assumed that the source and the receiver use the same basis vectors
for the representation and the measurement of the states.
This assumption is necessary to establish the baseline of the game between
the source and the receiver.

For the mixed state, the entropy is reasonably given by the von Neumann
value.
As mentioned before, the von Neumann entropy for a pure
state is zero. But an unknown pure state will
communicate real information to the receiver, indicating that the
von Neumann entropy is not a reasonable measure in this case.

We propose that $S_i $ represent the informational entropy of the quantum
system with the density matrix $\rho$:

\begin{equation}
 S_i(\rho) = - \sum_i \rho_{ii}  \log \rho_{ii}.
\end{equation}

This represents the average uncertainty that the receiver has in relation to
the quantum state {\it for each
measurement}. Should the manner of the preparation of the pure state
be known to the observer, he can choose a basis state function that would
completely describe it, and there would indeed be no information 
associated with it.

By appropriately adjusting the basis vectors, the 
receiver can change the value of this entropy.
The value of $S_i(\rho)$ is not a measure of the entropy at
the transmitting end.
It is the amount of entropy of the quantum system
that is accessible to the receiver.

Some properties of $S_i$ are:

\begin{enumerate}
\item $S_i (\rho) \geq S_n (\rho)$, and the two are equal only when
the density matrix has only diagonal terms.
\item $S_n (\rho)$ is obtained by minimizing $S_i (\rho)$
with respect to all possible unitary transformations.
In other words,
\begin{equation}
S_n (\rho) = \inf_{U} S_i (U\rho U^{\dag})
\end{equation}
\item The maximum value of $S_i $ is infinity, true for the case where
the number of components is infinite.
\end{enumerate}

From the point of view of the source, a finite system can also carry 
infinite information.
Let us now, for convenience, assume that the quantum state is coded in the
polarization of photons:

\begin{equation}
|\phi\rangle = \alpha |0\rangle + \beta |1\rangle,
\end{equation}

\noindent
where the states $|0\rangle$ and $|1\rangle$ represent horizontally and 
vertically
polarized photons, respectively, and $\alpha$ is real.
The information exchange protocol may be defined by the transmission,
according to a clock, of photons, which are detected
using appropriate circuits and polarizing filters. 
The task of the receiver is to estimate the value of $\alpha$
(and, implicitly, $\beta$).
The value of $\alpha$ could be written down as a decimal sequence
in a string of 0s and 1s, that represents a secret.

As far as the receiver is concerned, only one bit
of information is obtained from  any single photon.
On the other hand, since a large number of identically 
prepared photons is
available, one could hope to find the exact probability amplitude
values 
$\alpha$ and $\beta$
by testing out different hypotheses related to
the nature of the state function. 
To determine these values with any precision, testing of 
a large number of the photons is required so as to approach
ever closer the true, unknown value. 

The measurement could use a transformation, so that the
transformed photon is rotated to the $|0\rangle$ state. Thus,
the receiver needs a procedure where the measured values would let him
find the transformation matrix:

\begin{equation}
   G=\left[ \begin{array}{cc}
                              \alpha  & \beta^* \\
                              -\beta & \alpha \\
                               \end{array} \right]
\end{equation}

\noindent

If each test is
assumed to provide one bit of information, then such a specifically
prepared photon carries information determined by the precision 
available to the receiver to distinguish between different 
component states.

For further simplicity, we might consider the qubit to be defined
such that both the values of $\alpha$ and $\beta$ are real,
and $|\phi\rangle = cos \theta |0\rangle + sin \theta |1\rangle$.
We can speak
of the unknown state to be associated with the angle $\theta$ as 
follows:

\begin{equation}
                     G^+ =   \left[ \begin{array}{cc}
                                  cos \theta  & sin \theta \\
                                  sin \theta & -cos \theta \\
                               \end{array} \right]
\end{equation}

It appears that there is no efficient deterministic algorithm
to estimate $G^+$.

\vspace{0.2in}
\noindent
{\bf Conjecture:} There is no deterministic algorithm
that will identify $G^+$ in $O (n^k)$ steps, where $n$ is the
number of quantization levels of $\theta$ that can be distinguished by
the receiver.

At worst this problem belongs to the {\bf NP} class,
because if an oracle were to guess the correct $G^+$, it is
easy to check it, since applying this transformation
the photons will be transformed to the state $|0\rangle$.

\subsubsection*{A cryptographic context}
The above scenario may be viewed in the context of cryptography as follows.
Alice and Bob agree to use
a $n$-qubit long sequence of photons
with varying polarization angles
that represents their shared signature, which is unknown to the
eavesdropper.
The sent message can be signed by each with this unique signature
sequence that follows
the data sequence, and since the recipient knows what to expect, it can be
validated.

Since
the eavesdropper, Eve, 
cannot use actual polarization angles (the probability of
getting that correct being infinitesimally small), 
she can match the projections of the signature 
bits with her own guessed sequence of 0s and 1s.
She
has a probability of $2^{-n}$ of guessing the
projection 
of the sequence along specific
basis vectors. 

Note that even if her guessed sequence
turned out to be correct, it is unlikely to
work at future times, since the
polarization angles associated with the qubit sequence are unknown to her,
and their projection to any basis states chosen by her are
going to vary from trial to trial.

Clearly the information associated with each qubit in this setting
is infinite.
It is incorrect, therefore, to assign finite entropy to a system in
the case of maximal ignorance even if the system is finite.

\section{Properties of informational entropy}

Consider that the quantum system is represented by the density operator
$\rho$, which is an ensemble of pure states $|\phi_i\rangle$
with probabilities $p_i$ and a mixed state with density operator
$\rho_o$ with probability $p_o$
in the following manner:

\begin{equation}
\rho = \sum_i p_i |\phi_i \rangle \langle\phi_i| + p_o \rho_o.
\end{equation}

The total informational entropy of the system will be given by:

\begin{equation}
S_i (\rho) \geq \sum_i  p_i S_p (\phi_i) + p_o  S_n (\rho_o)
\end{equation}

\noindent
where $S_p (\phi)$ represents the entropy of the pure state 
$|\phi\rangle = \sum_k c_k |a_k\rangle$:
\begin{equation}
 S_p (\phi) = - \sum_k |c_k|^2 \log |c_k|^2 
\end{equation}
\noindent
that is a companion to the mixed state.
The reason why the left hand side can be larger than the sum of
the individual parts is that if the pure components are chosen
inappropriately, as aligned with the basis components at the receiver,
one would obtain no contribution towards entropy from such components.

\vspace{0.2in}
\noindent
{\bf Example 1.} Let the system density operator be described by:

\begin{equation}
\rho  =   \left[ \begin{array}{cc}
                                  .5  & .25 \\
                                  .25 & .5 \\
                               \end{array} \right]
\end{equation}

Here, 

\begin{equation}
\rho  = 0.5 \times  \left[ \begin{array}{cc}
                                  .5  & .5 \\
                                  .5 & .5 \\
                               \end{array} \right]
+ 0.5 \times  \left[ \begin{array}{cc}
                                  .5  & 0 \\
                                  0 & .5 \\
                               \end{array} \right]
\end{equation}

As far as the receiver is concerned, there is no way for him to
know {\it a priori} whether the quantum state received is pure
or mixed. In each test of the very many copies of the state available
to him (assumed in our communication protocol), he receives one
bit of information. The informational entropy in the beginning is
1 bit.

A simple calculation tells us that $S_i (\rho) = 1 $ bit, whereas 
$S_n (\rho) = 0.811$ bit. On the other hand, $S_p$ 
is 1 bit, and $S_i = 0.5 \times 1 + 0.5 \times 1 = 1 $ bit.
Clearly, informational entropy is a 
better measure than the von Neumann measure in this situation.

\vspace{0.2in}
\noindent
{\bf Example 2.} Let the system density operator be described by:

\begin{equation}
\rho  =   \left[ \begin{array}{cc}
                                  .71  & .15 \\
                                  .15 & .29 \\
                               \end{array} \right]
\end{equation}

The informational entropy for this example is -.71 log .71 - .29 log .29
= 0.868 bits.
\noindent
But this quantum state may be written down as the statistical ensemble:

\begin{equation}
\rho  = 0.3 \times  \left[ \begin{array}{cc}
                                  .5  & .5 \\
                                  .5 & .5 \\
                               \end{array} \right]
+ 0.7 \times  \left[ \begin{array}{cc}
                                  .8  & 0 \\
                                  0 & .2 \\
                               \end{array} \right]
\end{equation}

The first part of the ensemble represents a pure state with
a probability of 0.3 and the second part is a mixed state with
a probability of 0.7. One can easily calculate that the individual
components have information of 1 bit and 0.722 bits, respectively.
The sum of the two entropies is therefore:

\begin{equation}
0.3 \times 1 + 0.7 \times .722 = 0.805
\end{equation}

\noindent
bits, which is less than the informational entropy.

\vspace{0.2in}
\noindent
{\bf Example 3.} Let the system density operator be described by:

\begin{equation}
\rho  =   \left[ \begin{array}{cc}
                                  .75  & 0 \\
                                  0 & .25 \\
                               \end{array} \right]
\end{equation}

\noindent
The value of $S_i (\rho) = 0.559$ bits.

This system can also be expressed as a statistical ensemble with
two pure state components:

\begin{equation}
|a\rangle = \sqrt{\frac{3}{4}} |0\rangle + \sqrt{\frac{1}{4}} |1\rangle
\end{equation}

\begin{equation}
|b\rangle = \sqrt{\frac{3}{4}} |0\rangle - \sqrt{\frac{1}{4}} |1\rangle
\end{equation}

\noindent
and 

\begin{equation}
\rho = \frac{1}{2} |a\rangle \langle a| + \frac{1}{2} |b\rangle \langle b|
\end{equation}

The computation of the total entropy for this case is then:
\begin{equation}
S_i = \frac{1}{2} S_i (a) + \frac{1}{2} S_i (b)
\end{equation}

\noindent
Using the value of $S_i$ for each of the components, we get:

\begin{equation}
S_i = \frac{1}{2} \times 0.559  + \frac{1}{2} \times 0.559  = 0.559.
\end{equation}

This is exactly equal to the earlier calculation.
Or using two different ensembles of quantum states corresponding to
the same density matrix gives us identical results upon the use of
the informational entropy measure $S_i$.

\section{The origin of information}
Suppose the universe initially was in a pure or a low-entropy quantum state, 
how did
high entropy states arise? 
A part of the increase of entropy is due to the second law of
thermodynamics, but this contributes a small share to the
overall value.
Likewise, the expansion of the universe will contribute to the
increase, but this also does not square up with the actual
increase that has occurred.

Quantum evolution of a pure state leaves it unchanged and, therefore,
that cannot be the explanation for it.

But if we were to consider many particles in a pure state, say 
$|\phi\rangle  = \alpha |0\rangle + \beta |1\rangle$, then their
sequential observation will create mixed states of
non-zero von Neumann entropy.

Given the fact that we have both unitary, $U$, and non-unitary, $M_i$,
or measurement, operators, the density operator for each elementary 
state will change either to:

\begin{equation}
|\phi\rangle_{new} = \left\{ \begin{array}{ll}
		U |\phi\rangle & \mbox{unitary evolution}\\
		\frac{M_i |\phi \rangle }{\sqrt{\langle \phi | M_i^{\dagger} M_i | \phi\rangle }} & \mbox{non-unitary evolution}
	\end{array}
\right.
\end{equation}

When only non-unitary operators are used for the evolution, the 
elementary state will change from the pure state 
$|\phi\rangle  = \alpha |0\rangle + \beta |1\rangle$ to the 
mixed state given by the density matrix:

\begin{equation}
\rho  =   \left[ \begin{array}{cc}
                                  |\alpha|^2  & 0 \\
                                  0 & |\beta |^2 \\
                               \end{array} \right]
\end{equation}

\noindent
Its informational entropy would then have transformed completely
from that of the pure state to that of the mixed state, and its
von Neumann entropy would now be finite.

The existence of non-unitary operators requires the presence of
low-entropy structures that in themselves could not have arisen
in a universe governed by a single law. If gravitation is viewed
as the force that causes matter to aggregate, making non-unitary
evolution possible, then gravitation and quantum theory would
for ever be irreconcilable.

This view of the problem of how information increases 
is to postulate non-unitary evolution as a part of the earliest
universe [6], suggesting that unification has its limits.

\section{Complementarity and interaction-free measurement} 
We now consider information in the framework of distinguishing between
two states of an experimental arrangement that has traditionally been
associated with {\it interaction-free measurement} (IFM). Our intention is
to check the usefulness of the informational entropy measures in
this situation and to show that complementarity provides the
most reasonable explanation.

There are several versions of IFM, which are basically variants of
the Young's double-slit experiment, although for 
convenience the setting is
the Mach-Zehnder interferometer (Figure 1).
The basic idea of each is to focus on the counter-intuitive
fact that when the experiment is so set up that it is
possible to determine which path the photon took, the
photons exhibit particle-like behavior, and if it is not
possible to do so, then they exhibit wave-like behavior.

In Figure 1, a photon (from a source of single photons)
reaches the first half-silvered mirror, A (beam splitter)
 which has a transmission
coefficient 1. The transmitted and reflected parts of the
photon wave 
reunite at another, similar half-silvered mirror at D.
The beam splitters and fully-silvered mirrors (B and C) are arranged
in such a way that 
the photon is always detected by $D_1$, and 
never detected by $D_2$.
This corresponds to the baseline case where the entropy is zero, which
is reasonable given that there is no uncertainty associated with
the process.

\begin{figure}
\hspace*{0.2in}\centering{
\psfig{file=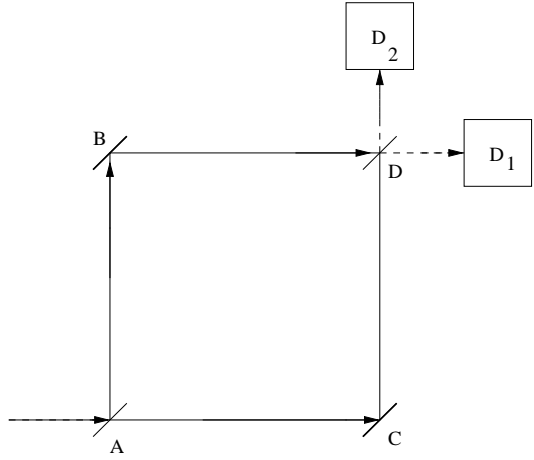,width=7cm}}
\caption{The Mach-Zehnder Interferometer}
\end{figure}

The IFM setting is associated with a modification to the
Mach-Zehnder interferometer by the use of a springy mirror
$C^+$ in place of $C$ as in Figure 2.
When the mirror C is rigid,
the photons will exhibit wave
nature; when the mirror is not rigid, the photons will exhibit
particle nature.
Figures 1 and 2 represent pure and mixed states, respectively.

There are three possible outcomes of this system:

\begin{description}
\item
i) photon absorbed, probability 1/2
\item
ii) detector $D_1$ clicks, probability 1/4
\item
iii) detector $D_2$ clicks, probability 1/4.
\end{description}

The difference between the two cases of Figure 1 and 2 is that
of the difference between a known pure state and a mixed state
with probabilities that are one-half in each of the component states.
It is appropriate to get the entropy in terms of the clicks.

In Figure 2, there is an equal probability that
the lower or the upper paths will be chosen. The choice of
the lower path leads to the absorption of the photon, whereas the choice of the
upper path leads to equal probability that it will end up
in $D_1$ or $D_2$.
If detector $D_2$ clicks,
one can claim that the photon was ``aware" 
that the mirror C was not rigid and it took the
upper path, and 
the spring in $C^+$ did not have to respond to the
photon. 
This arrangement corresponds to an entropy of 1.5 bits.

The claim is [7] that the method makes it possible to sense
an infinitely sensitive
mirror without interacting it with a probability
of 1/4. 
In reality, no measurement was necessary since we already knew that
the mirror $C^+$ is springy.

\begin{figure}
\hspace*{0.2in}\centering{
\psfig{file=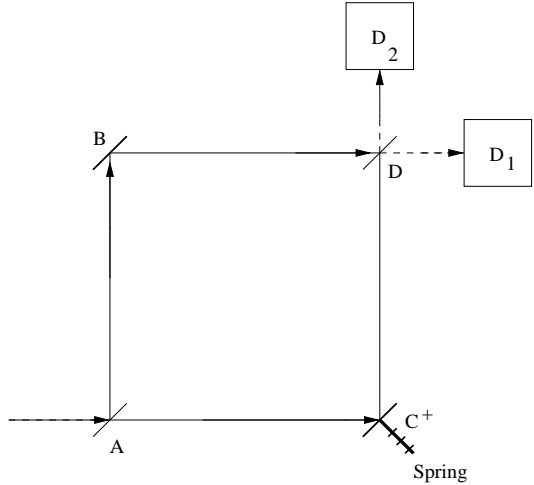,width=7cm}}
\caption{Mach-Zehnder interferometer with the mirror $C^+$ attached to a spring}
\end{figure}

If we wish to obtain information, it is essential that there be
alternatives. If we take it that we don't know if the experimental
arrangement consists of C or $C^+$ (both of which occur with equal
probability), then the outcomes are:

\begin{description}
\item
i) photon absorbed, probability 1/4
\item
ii) detector $D_1$ clicks, probability 5/8
\item
iii) detector $D_2$ clicks, probability 1/8.
\end{description}

Since this case is a mixture of the previous two cases, the entropy
will should be intermediate to the previous values. A simple calculation
gives the value as 1.299 bits.

Using Bayes' theorem, we know that

\begin{equation}
p(C^+ |D_2) = \frac{p(D_2|C^+) p(C^+)}{p(D_2)} = 1 
\end{equation}
\begin{equation}
p(C^+ |D_1) = \frac{p(D_1|C^+) p(C^+)}{p(D_1)} = 1/5 
\end{equation}

It is clear that it is the geometry of the experimental arrangement
that maps to different probabilities as listed above.
The alternatives of C and $C^+$ lead to pure and mixed
states, but the entropy associated with each of them is the same.

In the Many Worlds Interpretation (MWI), which is favoured 
by those who accept the reality of ``interaction-free measurement,''
one
assumes several worlds existing at the same time, and
in each world we perceive what is a small part of what is
in the universe. The laws of physics relate to the whole
universe, but viewed in the partial description of any
specific world, 
one may have paradoxical situations such as that of
measurement without interaction. 
In the framework of the MWI we find the springy mirror 
because
in another world it was indeed examined by a photon.

In the Complementarity Interpretation, one must speak
of the entire experimental arrangement.
The arrangement guarantees that the measurement is made,
albeit indirectly.

For example, the placement of C required prior measurement.
If the measurement consists of choosing between the arrangements
C and $C^+$, the placement of $C^+$ is associated with an uncertainty
due to the fact that one doesn't know in advance whether the mirror
will absorb the photon or not. Since

\begin{equation}
\Delta x \Delta p \geq \hbar /2 ,
\end{equation}

\noindent
and $\Delta p = h /\lambda$, therefore,

\begin{equation}
 \Delta x \geq \lambda / 4 \pi ,
\end{equation}

\noindent
where $\lambda$ is the wavelength associated with the photon.
If the location of $C^+$ cannot be precise, there will correspondingly
be uncertainty in the ability to distinguish between C and $C^+$.

\section{Non-unitary evolution}
Although non-unitarity is complementary to unitarity, it
may be seen as being generated by the measurement
process alone.
Let us consider the case where the state is evolving with time.
Let $|\psi_o\rangle$ be the initial state of quantum system, and let
the state evolve into $|\psi_t\rangle$ in time $t$. Let the
Hamiltonian characterizing the evolution be time-independent.

\begin{equation}
 |\psi_t\rangle = exp (- \frac{i}{\hbar} H t) |\psi_o\rangle
\end{equation}

Because of the continuing evolution of the state, any entropy
computation based on the von Neumann or the proposed pure state entropy
measure will be of fleeting significance. It appears, therefore, that
the entropy should be related to the unknown Hamiltonian $H$.

A measure of this entropy would be the frequency with which one
needs to observe the system so as to freeze the state, which brings
us to the so-called Zeno effect [8].
We can represent the evolution of the state by the following approximation:

\begin{equation}
|\psi_t\rangle  \approx (1 - \frac{i}{\hbar} H t - \frac{1}{2 \hbar^2} H^2 t^2) |\psi_o\rangle
\end{equation}

The correlation between the states at time  0 and t is:

\[ \langle\psi_o|\psi_t\rangle \approx \langle\psi_o|\psi_o\rangle -\frac{it}{\hbar} \langle\psi_o|H|\psi_o\rangle - \frac{t^2}{2 \hbar^2} \langle\psi_o|H^2|\psi_o\rangle \]
\begin{equation}
= 1  -\frac{it}{\hbar} \langle\psi_o|H|\psi_o\rangle - \frac{t^2}{2 \hbar^2} \langle\psi_o|H^2|\psi_o\rangle 
\end{equation}

\[ |\langle\psi_o|\psi_t\rangle|^2 \approx (1  -\frac{t^2}{2 \hbar^2} \langle\psi_o|H^2|\psi_o\rangle)^2 + \frac{t^2}{\hbar^2} \langle\psi_o|H|\psi_o\rangle^2 \]

\begin{equation}
= 1  -\frac{t^2}{\hbar^2} \langle\psi_o|H^2|\psi_o\rangle + \frac{t^2}{\hbar^2} \langle\psi_o|H|\psi_o\rangle^2 
\end{equation}

Let $ (\Delta E)^2 = \langle\psi_o|H^2|\psi_o\rangle - \langle\psi_o|H|\psi_o\rangle^2 $, then

\begin{equation}
|\langle\psi_o|\psi_t\rangle|^2  \approx 1 - \frac{(\Delta E)^2}{\hbar^2} t^2
\end{equation}

The evolution suppressing, Zeno case corresponds to repeated observations
at times $t/n$:

\[ |\langle\psi_o|\psi_t\rangle|^2  \approx (1 - \frac{(\Delta E)^2}{\hbar^2} \frac{t^2 }{n^2} )^n\]

\begin{equation}
= 1 - \frac{(\Delta E)^2}{\hbar^2} \frac{t^2}{n}
\end{equation}

We know that as the number of observations becomes infinite,
the state at time $t$ is the same as the state at time
$t=0$:

\begin{equation}
 \lim_{n \rightarrow \infty} \langle\psi_o|\psi_t\rangle|^2  \approx 1
\end{equation}

In our case, the measure of entropy would be the value of $n$ that
allows us to freeze the state within the precision available to the
receiver.

Consider a photon that is horizontally polarized, which we represent
by $|0\rangle$. We can, by using a polarizing filter, oriented
in the direction $45^o$, make half the number of photons collapse
to the state $\frac{1}{2} (|0\rangle + |1\rangle)$. In two
such observations, the photon's polarization would be steered to
$90^o$ with a probability of $\frac{1}{4}$ .

If the rotation in each step is $\theta^o$, one would need a total
of $\frac{\pi}{2 \theta} = n$ steps to rotate the original state of
$|0\rangle$ to the state of $|1\rangle$, and this will happen with the
probability of

\begin{equation}
(cos^2 \theta )^{\frac{\pi}{2 \theta}} 
\end{equation}

\begin{figure}
\hspace*{0.2in}\centering{
\psfig{file=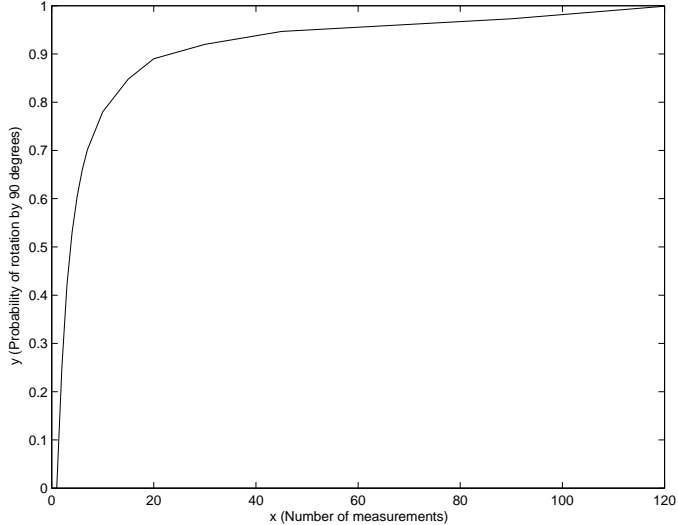,width=9cm}}
\caption{Observation driven evolution}
\end{figure}

Figure 3 illustrates this and
the probability of steering the photon to the
desired final state of $|1\rangle$ become quite close to 1 as $n$
approaches 100. For n =90, the probability is 0.973.

For someone who did not know that the photon was being steered by
repeated measurements, the evolution of the photon would be viewed as
a consequence of the Hamiltonian associated with the system.
If the measurements are made at regular intervals, the
Hamiltonian would be considered time-dependent.
The rotation would be largest at the first step and it will progressively
decrease with each new step.
Alternatively, one may view the rotation process to be faster
(associated with larger energy) at first with the speed tapering off
as observations continue.

If it is valid to see non-unitarity as resulting from wave collapse
alone, then 
the search for hidden variable theories of quantum mechanics will be
futile.
In this view, complete unification is not possible.

\section{Concluding Remarks}
Considering the information transfer problem from the point of
view of the preparer of the state and the experimenter, it is clear
that both mixed and pure states provide information to the
experimenter. For a two-component elementary mixed state, the most
information in each measurement is one bit, and each further 
measurement of identically prepared states will also be one bit.

For an unknown pure state, the information in it represents the choice
the source has made out of the infinity of choices related to
the values of the probability amplitudes with respect
to the basis components of the receiver's measurement
apparatus. The maximum information
in a pure state is thus infinite. On the other hand, each measurement
of a two-component pure state can provide one bit of information. But
if it is assumed that the source has made available an unlimited 
number of identically prepared states, the receiver can obtain
additional information from each measurement until the probability
amplitudes have been correctly estimated. Once that has occurred,
unlike the case of a mixed state, no further information will be
obtained from testing additional copies of this pure state.

The receiver can do this by adjusting the basis vectors so that he gets
{\it closer} to the unknown pure state. As the adjustment proceeds,
the amount of information that he would obtain from each measurement will
decrease. 
The information that can be obtained from such a state in
repeated experiments is 
potentially infinite in the most general case. 

But if the observer is told what the pure state is, the information
associated with the states vanishes, suggesting that a fundamental
divide exists between objective and subjective information.

The analysis of this paper is consistent with the positivist
view that one cannot speak of information associated with a
system excepting in relation to an experimental arrangement together
with the protocol
for measurement. The experimental arrangement is thus integral to
the amount of information that can be obtained.

The informational measure proposed in this paper resolves the puzzle of
entropy increase in the universe.
We can suppose that the universe had immensely large
informational entropy in the beginning, a portion of which
has, during the physical evolution of the universe,
transformed into thermodynamic entropy. 
If we take it that the dichotomy 
of quantum processes and gravitation is responsible for
unitary and non-unitary evolution, then it should not be
possible to unify the two.

The process of scientific discovery in terms of
the knowledge of its laws may be taken to be the unveiling of the
basis vectors of the pure state associated
with the bulk of informational entropy.
This process will be unending,
even as we come ever closer to the essential bases.

\end{document}